\documentclass[aps,prb,groupedaddress,twocolumn,10pt,showpacs]{revtex4-1}
\usepackage[pdftex]{graphicx}
\usepackage{amsmath,amssymb,color,subfigure}
\usepackage{amsthm}
\usepackage{amssymb}
\usepackage{braket}
\usepackage{mathrsfs}
\usepackage{setspace}

\newcommand{\be}{\begin{equation}}
\newcommand{\ee}{\end{equation}}
\newcommand{\bea}{\begin{eqnarray}}
\newcommand{\eea}{\end{eqnarray}}

\begin{document}
\title{Tuning the Chern number and Berry curvature  \\
with spin-orbit coupling and magnetic textures}
\author{Timothy M. McCormick}
\author{Nandini Trivedi}
\affiliation{Department of Physics, The Ohio State University, Columbus, OH  43210, USA}
\date{%This version started 2014-11-19; compiled 
\today}

\begin{abstract}
We obtain the band structure of a particle moving in a magnetic spin texture, classified by its chirality and structure factor, in the presence of spin-orbit coupling.
This rich interplay leads to 
a variety of novel topological phases characterized by the Berry curvature and their associated Chern numbers.
We suggest methods of experimentally exploring these topological phases by Hall drift measurements of the Chern number and 
Berry phase interferometry to map the Berry curvature.

\end{abstract}
%\pacs{}
\maketitle

%\tableofcontents
\section{Introduction}

Strong spin-orbit coupling lies at the heart of a variety of novel electronic phenomena, including topological band insulators (TBI) \cite{hasanKane,qiZhang}
and Weyl semimetals \cite{weyl}.  In a TBI the electronic band structure
is topologically distinct from that of a trivial insulator.  The presence of zero energy edge modes,
similar to the chiral boundary states of the quantum Hall effect, is a direct consequence of the topological order which defines a TBI.  The prediction 
\cite{kaneMeleGraph,BHZ,fuKaneMele,mooreBalents} and subsequent experimental verification \cite{molenkampExp,hsiehEtAl,zhangVer}
of novel topological phases in a variety of materials has spawned a vast new field of topological quantum matter.

Recent developments in ultracold atomic gases allow spin-orbit coupling to be tuned using Raman processes \cite{socCold1,socCold2,socCold3},
expanding this frontier beyond electronic systems.  Cold atom systems are already an excellent venue for the exploration of many-body physics \cite{blochDalRev}
and the ability to simulate spin-orbit coupling promises to allow a detailed study of the interplay of spin-orbit coupling and strong interactions.  
Recent theoretical predictions \cite{cole2012,radicGalitski} suggest that, on a lattice,
the Bose-Hubbard model with spin-orbit coupling gives rise to a rich collection of effective magnetic Hamiltonians
in the strongly interacting limit that support a plethora of novel magnetic states, such as ferromagnets, 
antiferromagnets, spirals, and chiral textures.

It is well known that systems with broken time-reversal symmetry exhibit off-diagonal Hall conductivity \cite{hallPap}, and this transverse conductivity is nonzero 
in magnetic materials even in the absence of an external magnetic field \cite{hallAHE}. The study of the anomalous Hall effect has been extended to a quantum
geometrical viewpoint, with the Berry phase of the wavefunction taking a central role in this phenomena.  
The quantum anomalous Hall effect (QAH), a quantized version of the anomalous Hall effect, exhibits edge states which carry a quantized transverse conductivity, similar
to the quantum Hall effect but without the requirement of an externally applied magnetic field.  The origin of the QAH effect 
lies in an exchange interaction between a spinful itinerant particle and localized magnetic moments rather than from Landau levels due to the orbital effects of an external magnetic field.
Although the QAH effect offers the promise of low dissipation transport without the need for an external magnetic field, there are few experimentally accessible systems which offer
the combination of topologically nontrivial insulating behavior and magnetic structure.  Despite these challenges, the QAH effect has been observed in thin films
of Cr-doped Bi$_2$Te$_3$, a magnetic topological insulator \cite{qahObs}.

In this paper we propose a method for realizing the QAH effect in cold atom systems featuring itinerant particles with spin-orbit coupling moving in a chiral spin texture.  We begin
with a tight-binding model with Rashba spin-orbit coupling and a Hund's coupling to a fixed magnetic texture.  After deriving an effective Hamiltonian for the system in the adiabatic
approximation where the itinerant spin is always parallel to the local texture, we investigate some consequences of this effective Hamiltonian for ferromagnetic, antiferromagnetic
spiral and chiral textures.  We show that chiral textures lead to a topologically nontrivial band structure for the itinerant particles.  By changing the strength of the spin-orbit coupling for the itinerant 
particle, we find the system undergoes a variety of topological phase transitions
between various quantum anomalous Hall states.  Finally, we propose that our results can be experimentally verified using Hall drift experiments and Berry phase interferometry.

\section{Model}

\subsection{Hamiltonian}

We consider a single particle with two internal degrees of freedom, denoted $ \uparrow $ and $ \downarrow $,
on a two dimensional square lattice interacting with a localized spin texture through Hund's coupling.  
We describe our system by the following Hamiltonian:
\begin{equation}
\label{parentHam}
H_{\textrm{full}} = -t \sum\limits_{\langle i,j \rangle} (\Psi_{i}^\dagger \mathcal{R}_{ij} \Psi_{j} + h.c.) - J \sum\limits_{i} \mathbf{S}_{i}\cdot 
c^{\dagger}_{ia} (\vec{\sigma}_{i})_{ab} c_{ib}, 
\end{equation}
where $\Psi_{i}^\dagger = (c_{i\uparrow}^\dagger , c_{i\downarrow}^\dagger)$ is a spinor of creation operators 
and the matrix $\mathcal{R}_{ij} \equiv \mathrm{exp}[i\mathbf{A}\cdot(\mathbf{r}_{i} - \mathbf{r}_{j})]$.
We see that on-diagonal elements of $ R_{ij} $ describe spin-preserving hopping, while a non-Abelian gauge field
$\mathbf{A} = (\alpha \sigma_y , \beta \sigma_x , 0)$ leads to off-diagonal elements which describe spin flip hopping.  
We set $\beta = - \alpha$ in order to obtain the lattice analog of Rashba spin-orbit coupling which we use for the remainder of this paper.  
We also take the lattice spacing to be unity such that for $i,j$ which are neighbors, $|\mathbf{r}_{i} - \mathbf{r}_{j}| = 1$.
The spin $\vec{\sigma}_{i}$ of the itinerant particle couples to the local spin texture denoted by $ \mathbf{S}_{i} $ with a coupling strength $ J > 0$.  

We assume that
the time scale of the itinerant particle's evolution is much shorter than that of the texture and so we consider the localized spins to be fixed in the 
spirit of the Born-Oppenheimer approximation.  This can be accomplished in a cold atom system by creating the texture with a heavy species of strongly 
interacting particles and allowing them to strongly couple with a lighter species of particles.  The single particle approximation can
be accomplished experimentally by taking the interactions between light particles to be exceedingly weak.

\subsection{Local Projection}

In the limit of $J/t \rightarrow \infty$, the coupling 
to the texture will split the spectrum into two bands and the low energy behavior of the system will be governed entirely by the lower band which is aligned locally
with the spin texture.  Because of this local alignment, we introduce operators representing the local orientation of the spin $f_{i\sigma}^{\dagger}$ which creates a spin
aligned with the local spin $\mathbf{S}_{i}$ at site $i$.  In the usual manner, we can write these in terms of the creation and annhilation 
operators for spins aligned with the global $z$-direction.  We obtain the following unitary transformation
\be
\label{locTrans}
  \left[ {\begin{array}{c}
   f_{j\uparrow}^{\dagger} \\
   f_{j\downarrow}^{\dagger} \\
  \end{array} } \right] =
  \left[ {\begin{array}{cc}
   g_c(\theta_j,\phi_j) & g_s(\theta_j,\phi_j) \\
   g^{*}_{s}(\theta_j,\phi_j) & -g^{*}_{c}(\theta_j,\phi_j) \\
  \end{array} } \right]%\cdot
  \left[ {\begin{array}{c}
   c_{j\uparrow}^{\dagger} \\
   c_{j\downarrow}^{\dagger} \\
  \end{array} } \right],
\ee where $g_c(\theta_j,\phi_j)$
gives the component of the projection parallel to the global axis and 
$g_s(\theta_j,\phi_j)$
gives the component of the projection antiparallel to the global axis.
The orientation of the local spin at a site $j$ is described by the polar and azimuthal angles $(\theta_j,\phi_j)$ with respect to the global $z$-axis.

With Eq. (\ref{locTrans}), we can write the Hamiltonian
in terms of the local spin creation and annihilation operators.  For $J/t \rightarrow \infty$, we can now project out the space where the spins are locally
aligned with the texture by keeping only the terms in the Hamiltonian
which are of the form $f_{i \uparrow}^{\dagger}f_{j \uparrow}$. 

The Hamiltonian now takes the form:
\be
\label{projHam}
H_{\textrm{proj}} = -\sum_{i,\vec{\delta}} \big( t_{i,i+\vec{\delta}}' \cos(\alpha) + t_{i,i+\vec{\delta}}'' \sin(\alpha) \big)
f_{i \uparrow}^{\dagger}f_{i+\vec{\delta} \uparrow}
\ee
where $\vec{\delta} \in \{ \pm \hat{x} , \pm \hat{y} \}$ is a nearest neighbor of the site $i$.  The hopping amplitudes $t_{i,i+\vec{\delta}}'$ 
and $t_{i,i+\vec{\delta}}''$ now encapsulate the information
about the spin texture, the spin-orbit coupling and the direction $\vec{\delta}$ of the nearest neighbor.
The explicit forms of the modified hopping matrix elements $t^\prime$ and $t^{\prime\prime}$ are given in Appendix A.

\begin{figure}[!t]
\centerline{\includegraphics[width= .46 \textwidth]{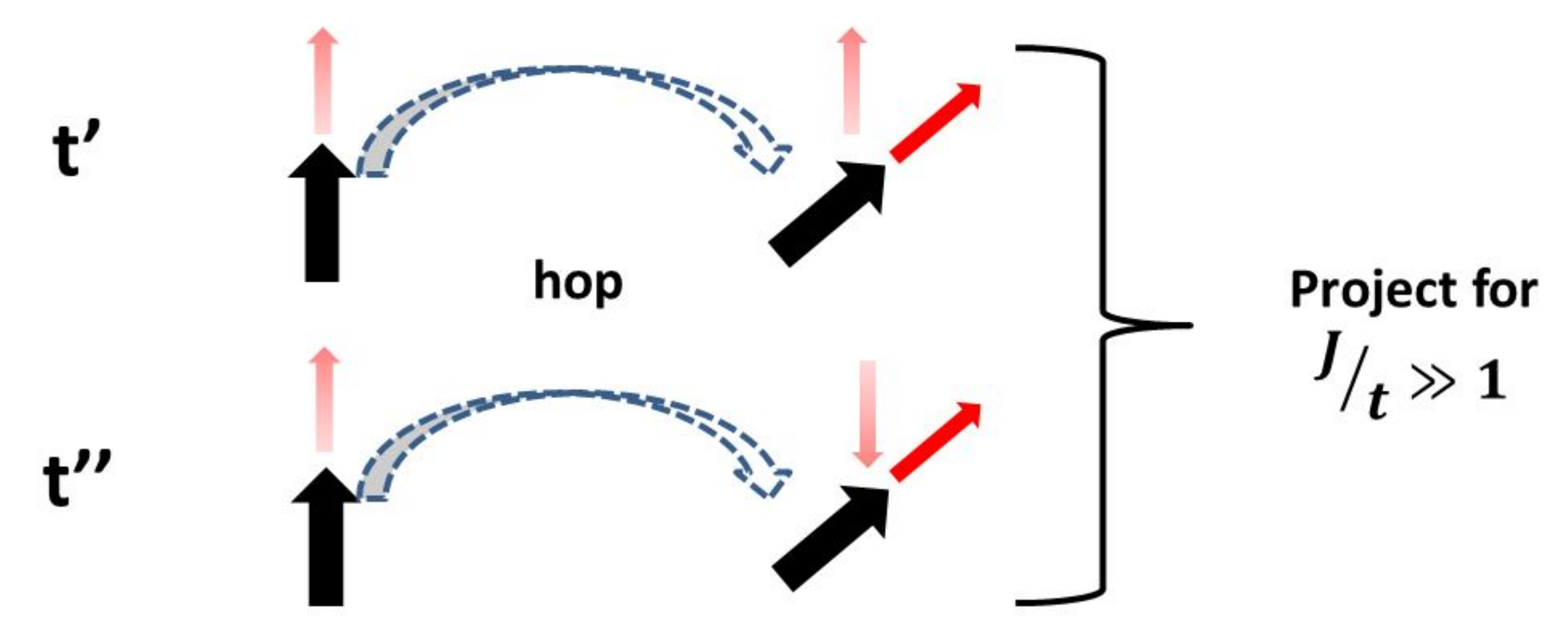}}
\caption{A schematic illustration of the hopping processes which lead to the matrix elements $t'$ and $t''$.  
Due to the nature of the spin-orbit coupling, the hopping amplitudes now depend on the direction of the hopping
as well as the orientation of the spin texture at each site.  The faded red arrows show how
a particle hops and then projects along the local texture and the dashed blue arrow represents the hopping.
Particles hop without changing their spin and then project onto the local 
texture for $t'$. For $t''$, particles flip their spin from spin-orbit coupling while they hop to adjacent sites and then align with the local texture. }
\label{projCartoon}
\end{figure}
Embedded in this Hamiltonian is a real space vector potential which arises from the combination of the spin texture and Rashba spin-orbit coupling.  
%In the limit of no spin-orbit coupling, this vector potential creates magnetic flux through each plaquette equal to the solid angle subtended by the spins at the corners of that plaquette.

\subsection{Characterization of Spin Textures}

\begin{figure*}[!t]
\centerline{\includegraphics[width= 1.17\textwidth]{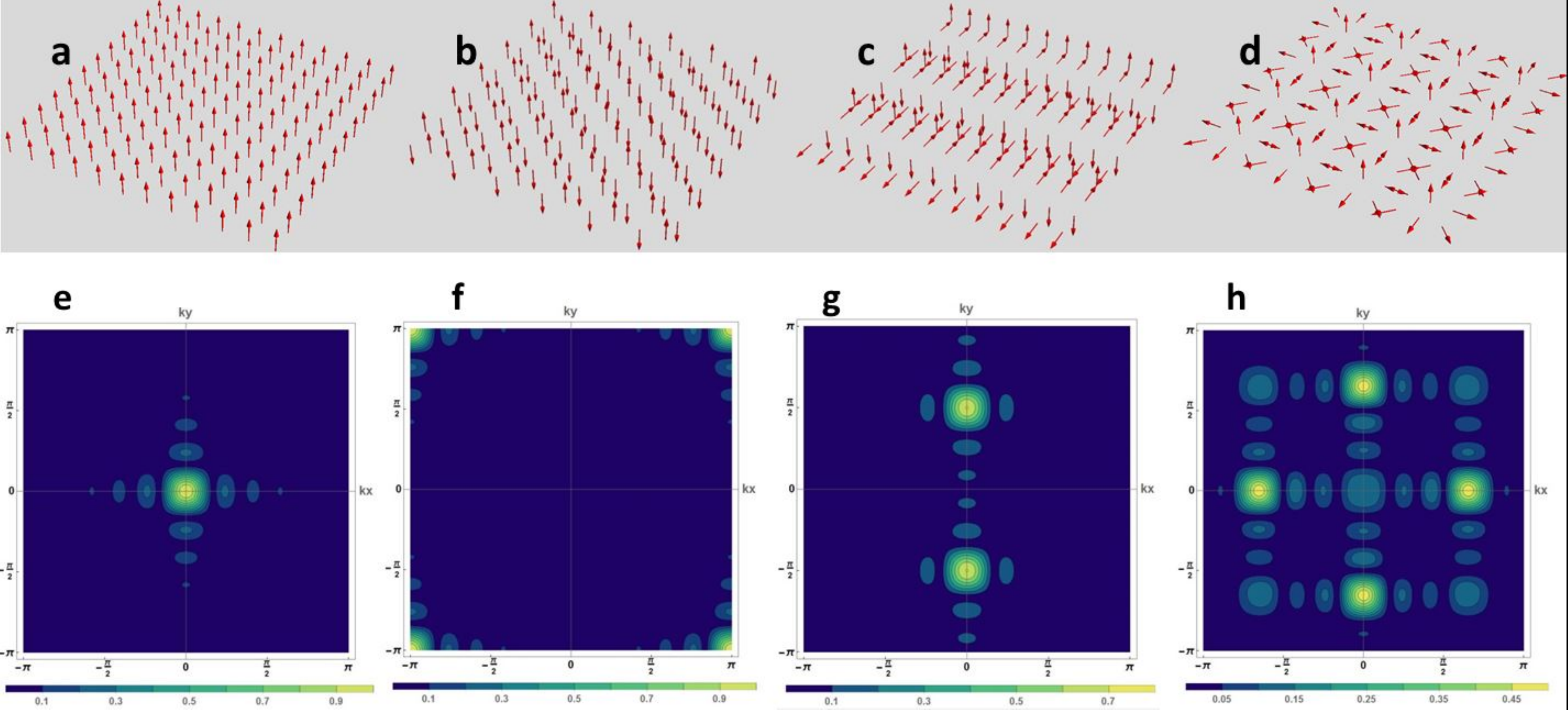}}
\caption{
(a-d)
The real space spin configurations for several spin textures: (a) a ferromagnetic texture, (b) an antiferromagnetic texture, (c) 4x1 spirals on a 12x12
lattice, (d) 3x3 merons on a 12x12 lattice. 
(f-h) The spin structure factors for several textures on a 12 by 12 site lattice: (e) a ferromagnetic texture, (f) an antiferromagnetic texture,
(g) spirals with a 4x1 unit cell in the $y$ direction, (h) sixteen 3x3 merons}
\label{textsAndStructs}
\end{figure*}

We fix a given spin texture a priori by choosing a set of spin configurations $\{ \theta_i, \phi_i  \}$.
In order to study the interplay of spin-orbit coupling and coupling to a real-space spin texture, we 
investigate the effects of ferromagnetic, antiferromagnetic, spiral and chiral spin textures.  Examples of some
of these textures are shown in Fig. \ref{textsAndStructs}(a-d).

These textures can be broadly characterized by their magnetic structure factor
\be
\label{magStruct}
S(\mathbf{k}) = | \sum_{j} e^{i \mathbf{k}\cdot \mathbf{r}_{j}} \mathbf{S}_j |,
\ee
which is simply the Fourier transform of the local spin texture.  In quantum materials it can be measured 
by neutron diffraction and in cold atom systems it can be accessed 
by Bragg spectroscopy\cite{braggHulet}.  
%The momenta $\mathbf{k}_{\textrm{max}}$ for which $S(\mathbf{k})$ is maximal correspond to the angles with which the texture varies between neighboring spins.
%To illustrate how the peaks of the magnetic structure factor can characterize the spin textures, 
We show $S(\mathbf{k})$ in Fig. \ref{textsAndStructs}(e-h) in the first Brillouin zone
for a variety of textures on a 12 site by 12 site lattice.

We also characterize a spin texture by its chirality.  On a lattice, we define the local scalar chirality at a site $i$ as 
\be
\label{locChir}
\chi_{i} = \dfrac{1}{8 \pi} \big( \mathbf{S}_i \cdot ( \mathbf{S}_{i+\hat{x}} \times \mathbf{S}_{i+\hat{y}} ) + \mathbf{S}_i \cdot ( \mathbf{S}_{i-\hat{x}} \times \mathbf{S}_{i-\hat{y}} ) \big).
\ee
We will refer to $ \chi = \sum_{i} \chi_{i} $ as simply the chirality for a given texture.
This quantity will obviously be zero for collinear or coplanar spin textures, such as ferromagnetic, antiferromagnetic and spiral textures.  
We call textures for which $ \chi \ne 0 $ as chiral textures.
The chirality is a topological invariant of the texture.

\textit{Ferromagnet}:  We consider ferromagnetic (Fig. \ref{textsAndStructs}a) textures oriented along the $z$ axis.  The magnetic structure factor exhibits a peak at $\mathbf{k} = [0,0]$
as shown in Fig. \ref{textsAndStructs}e.

\textit{Antiferromagnet}: In the antiferromagnetic phase (Fig. \ref{textsAndStructs}b), spins point alternately along the $\pm z$ axis.  
The magnetic structure factor has a maximum at $\mathbf{k} = [\pi,\pi]$ as shown in Fig. \ref{textsAndStructs}f.

\textit{Spiral}: Spiral phases are composed of spins which cant with constant angle in one direction and which do not vary in an orthogonal direction.  For simplicity, we consider spiral
textures for which the spiral is oriented in either the $x$ or $y$ directions, rather than at an arbitrary angle.  For a commensurate spiral oriented in the $x$ ($y$) direction that winds
with a period of $L$ sites, the structure factor has a peak at $[\pm 2\pi/L,0]$ ($[0,\pm 2\pi/L]$) 
as shown in Fig. \ref{textsAndStructs}c.  We call the angle $\psi = 2\pi/L$ between adjacent spins the canting angle.
We note that since spiral textures are coplanar, they have zero chirality.
A spiral texture with a 4x1 unit cell is shown in Fig. \ref{textsAndStructs}g.

\textit{Meron}: A non-coplanar texture in which the center spin in a unit cell is aligned with the $z$ axis and spins at the edge of the unit cell take a polar
angle of $\pi/2$ with respect to the $z$ axis. Mapping the orientation of spins in a unit cell to their location on the unit sphere, one finds that the sphere is half covered.
Merons have $ \chi = 1/2 $.  A superlattice of 3x3 merons is shown in Fig. \ref{textsAndStructs}d and its structure factor is shown in Fig. \ref{textsAndStructs}h.

\textit{Skyrmion}: Skyrmions are non-coplanar textures where the center spin in a unit cell is aligned with the $z$ axis and spins at the edge of the unit cell are aligned
with the $-z$ axis. Mapping the orientation of spins in a unit cell to their location on the unit sphere, one finds the sphere to be fully covered.
Skyrmions have $ \chi = 1 $ and they can be considered as a composite object composed of two merons.

\section{Results}

We have exactly diagonalized the above single particle Hamiltonian for several spin textures at a variety of spin-orbit couplings. After obtaining
the single particle spectrum, a wide variety of quantities can be investigated.
We derive effective Hamiltonians in the case of ferromagnetic, antiferromagnetic, spiral and meron textures. 
We then characterize the topological nature of the bulk band structure by calculating the Berry curvature and the Chern number for each band.

\subsection{Projected Hamiltonian in Simple Textures}

\begin{figure*}[!t]
\centerline{\includegraphics[width= 0.85 \textwidth]{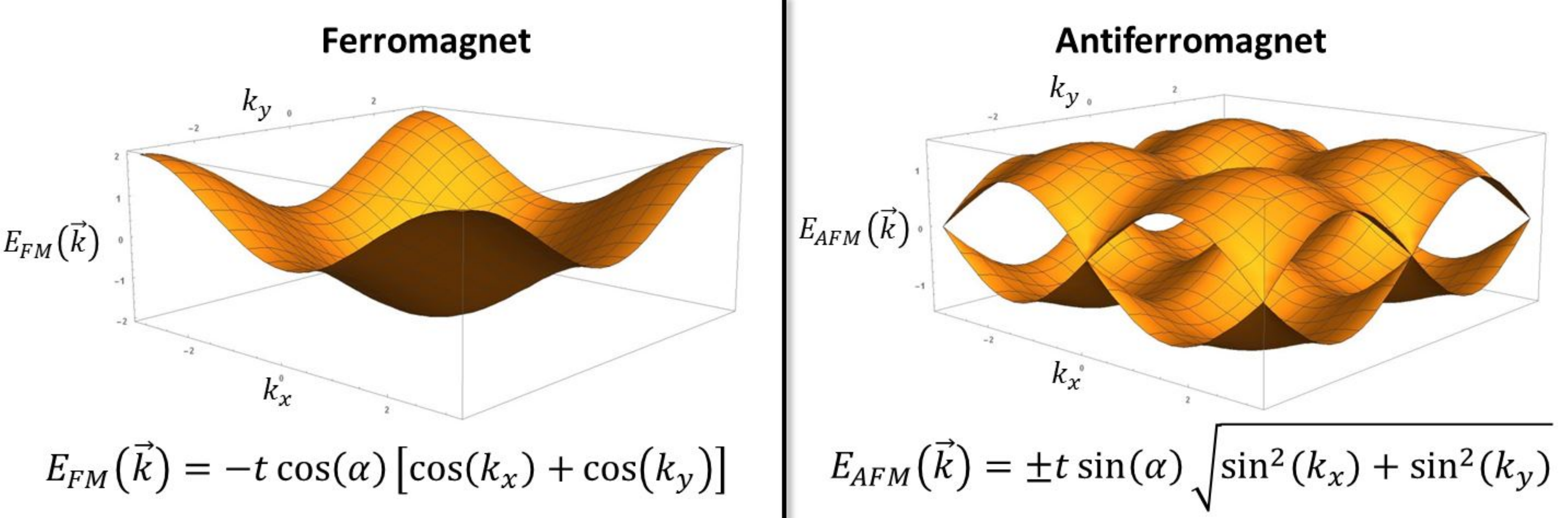}}
\caption{	
(Left) The band structure for a particle in a ferromagnetic texture is identical to that of a free particle with a $\cos (\alpha)$ modulated bandwidth.
(Right) The bands for an antiferromagnetic texture are identical to that of the Rashba portion of the Hamiltonian in Eq. (\ref{parentHam}) with spin-preserving hopping
with a $\sin (\alpha)$.
There are Dirac cones at the time-reversal invariant momenta for the antiferromagnetic bands.
}
\label{afmBands}
\end{figure*}

For a purely ferromagnetic texture,
$t_{i,i+\vec{\delta}}''$ is zero for all $i$ and $t_{i,i+\vec{\delta}}'$ is unity for all $i$.  It is clear that by projecting
into the locally aligned basis, systems with ferromagnetic textures are similar to a free particle for closed boundary conditions or the familiar particle in a box
for those with open boundary conditions.  Spin-orbit
coupling leads to a modulated effective hopping strength 
\be
E_{\textrm{FM}}(\mathbf{k}) = -t\cos(\alpha) [ \cos(k_x) + \cos(k_y) ]
\label{fmProj}
\ee
shown in the left panel of Fig. \ref{afmBands}.

Due to the opposite orientation of neighboring spins in the antiferromagnetic texture, it is clear that $t_{i,j}'$ is identically zero for all values of spin-orbit coupling $\alpha$.
Without spin-orbit coupling, this system is trivial, since the local projection destroys hopping amplitudes between oppositely aligned spins.  The spin-flip
terms in the Hamiltonian in Eq. (\ref{parentHam}) dominate the contribution to the hopping elements in projected hopping elements in
Eq. (\ref{projHam}) and lead to an effective Hamiltonian for the antiferromagnetic texture
\be
\label{afHam}
H_{\textrm{AF}} = -2 t\sin(\alpha)\sum_{\mathbf{k}} \Phi^{\dagger}_{\mathbf{k}\uparrow} \mathcal{H}_{\textrm{AF}}(\mathbf{k}) \Phi_{\mathbf{k}\uparrow}.
\ee
Here $\Phi^{\dagger}_{\mathbf{k}\uparrow} = (f^{\dagger}_{\mathbf{k}\uparrow A},f^{\dagger}_{\mathbf{k}\uparrow B})$ where the 
$f^{\dagger}_{\mathbf{k}\sigma}$'s are defined as the Fourier transform of the operators defined in Eq.(\ref{locTrans}) and the A or B refers to
the two sublattices of the antiferromagnetic texture.  The kernel $\mathcal{H}_{\textrm{AF}}(\mathbf{k})$ in Eq. (\ref{afHam}) is given by
\be
\label{afKernel}
\mathcal{H}_{\textrm{AF}}(\mathbf{k}) = \sin(k_x) \tau_x + \sin(k_y) \tau_y
%CHECK: I verified this - TM
\ee
where $\vec\tau$ are the pseudo spin Pauli matrices associated with the A and B sublattices.  We find the energy dispersion shown in the right panel of Fig. \ref{afmBands} to be 
\be
E_{\textrm{AF}}^{\pm}(\mathbf{k}) = \pm 2 t \sin(\alpha) \sqrt{\sin^{2}(k_x)+\sin^{2}(k_y)}.
\label{afmnrg}
\ee
This spectrum is identical to that of the Hamiltonian for a free particle with Rashba spin-orbit coupling and no spin-preserving hopping.  Here we find Dirac cones 
at the time-reversal invariant momenta.  We note that since we do not obtain a gapped band structure, we cannot define a topological invariant for the antiferromagnetic case.

\begin{figure*}[!t]
\centerline{\includegraphics[width= 1.0 \textwidth]{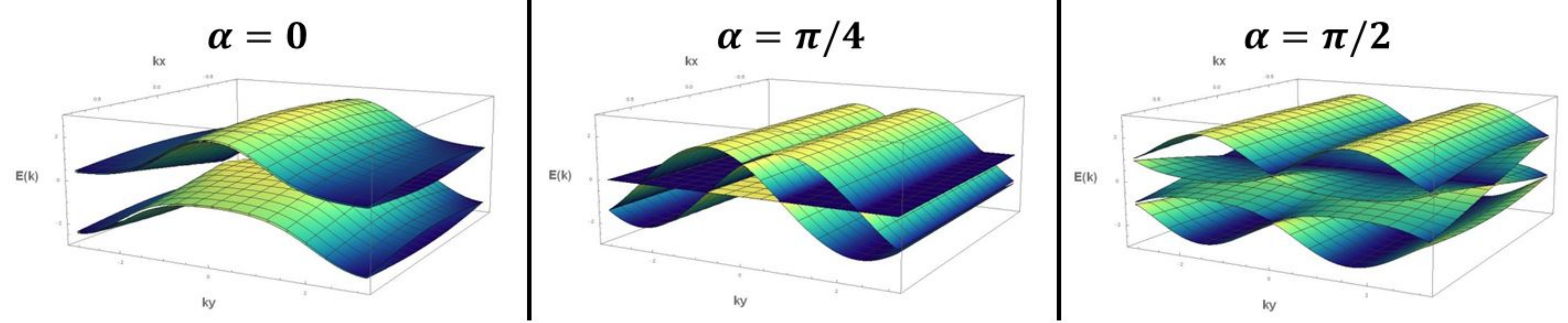}}
\caption{The band structure in the reduced Brillouin zone
for a 4x1 spiral texture with $\alpha = 0$ (left), $\alpha = \pi/4$ (middle) and $\alpha = \pi/2$ (right).  We note the presence of zero energy bands for $\alpha = \pi/4$.}
\label{spiralB}
\end{figure*}

For a spiral texture, we pick the direction of the spiral canting to be in the $x$-direction without loss of generality. 
Motivated by recent theoretical predictions \cite{cole2012}, we investigate specifically the case of $L = 4$ spiral. For this texture, the band structure does not have a simple closed analytic
form.  Notably, both $t'$ and $t''$ are non-zero, leading to a nontrivial dependence of the dispersion on $\alpha$.  In Fig. \ref{spiralB}, we show the band structures for
$\alpha = 0, \pi/4 \textrm{ and } \pi/2$ across the reduced Brillouin zone \{$-\pi \le k_{y} < \pi$, $-\pi/4 \le k_{x} < \pi/4$\}.  We draw particular attention to the flat bands at zero energy
for $\alpha = \pi/4$ where commensuration of spin-orbit coupling and the angle of the spiral lead to destructive interference which eliminates hopping along the direction of the spiral.

\subsection{Chern Numbers and Berry Curvature}

The band structure of a periodic system determined by a Bloch Hamiltonian $H(\mathbf{k})$ can be characterized by the Chern number 
\cite{TKNN,kohmotoCher,berryPhaseElec}.  
The Chern number for the \textit{n}$^{th}$ Bloch band is defined as
\be
\label{chernNumDef}
c_n = \dfrac{1}{2\pi} \int{d^2\mathbf{k}\cdot \nabla_{\mathbf{k}} \times \mathbf{A}_n(\mathbf{k})}
\ee
where the Berry connection $\mathbf{A}_n(\mathbf{k})$ is defined in terms of the wavefunction of the \textit{n}$^{th}$ band $\ket{n(\mathbf{k})}$ as
\cite{berryOrig}
\be
\label{berryConnDef}
\mathbf{A}_n(\mathbf{k}) = \bra{n(\mathbf{k})}i \nabla_{\mathbf{k}}\ket{n(\mathbf{k})}.
\ee
The Hall conductivity is given by
\be
\label{hallCond}
\sigma_{xy} = \dfrac{e^2}{h} \sum_{n} c_n,
\ee
where the summation index $n$ runs over the filled bands.

The presence of a spin texture in the Hamiltonian in Eq. (\ref{parentHam}) explicitly breaks time-reversal symmetry in general 
and so the Chern number $c_n$ of the \textit{n}$^{th}$ band of $H_{\textrm{proj}}$ in Eq. (\ref{projHam}) will not be zero in general.
A square spin texture with sides of length $L$ in a system with periodic boundary conditions is described by a Bloch Hamiltonian which can 
be represented as an $L^2$ dimensional matrix in momentum space.
Due to the numerical nature of our exact diagonalization, we can diagonalize our system only on a discrete mesh of points $k_l$ within the
Brillouin zone. 

Following Fukui, Hatsugai and Suzuki \cite{hatsugaiDiscChern}, we introduce a U(1) link variable 
\be
\label{u1link}
U_{n,\mu}(k_l)=\langle n(k_l)|n(k_l + \hat{\mu})\rangle / |\langle n(k_l)|n(k_l + \hat{\mu})\rangle|
\ee
and a lattice field $F$ analogous to the continuum Berry curvature $\nabla_{\mathbf{k}} \times \mathbf{A}_n(\mathbf{k})$:
\be 
\label{latField}
F_{n}(k_l) = \ln \bigg( \dfrac{U_{n,x}(k_l)U_{n,y}(k_l + \hat{x})}{U_{n,x}(k_l + \hat{y})U_{n,y}(k_l)} \bigg),
\ee
where we choose $F(k_l)$ to lie within the principal branch of the logarithm. In the logarithm in $F(k_l)$, the U(1)
gauge field is summed around a single plaquette starting at the momentum $k_l$.
Lastly, we take the lattice Chern number to be 
\be
\label{latChern}
\widetilde{c}_n = \dfrac{1}{2\pi i} \sum_{l} F_{n}(k_l)
\ee
where $l = 1,...,L^2$.  In addition to providing a robust method of calculating the Chern number for a modestly sized mesh
of points in momentum space, the lattice field $F_{n}$ is manifestly gauge invariant.

\begin{figure*}[!t]
\centerline{\includegraphics[width=1.0 \textwidth]{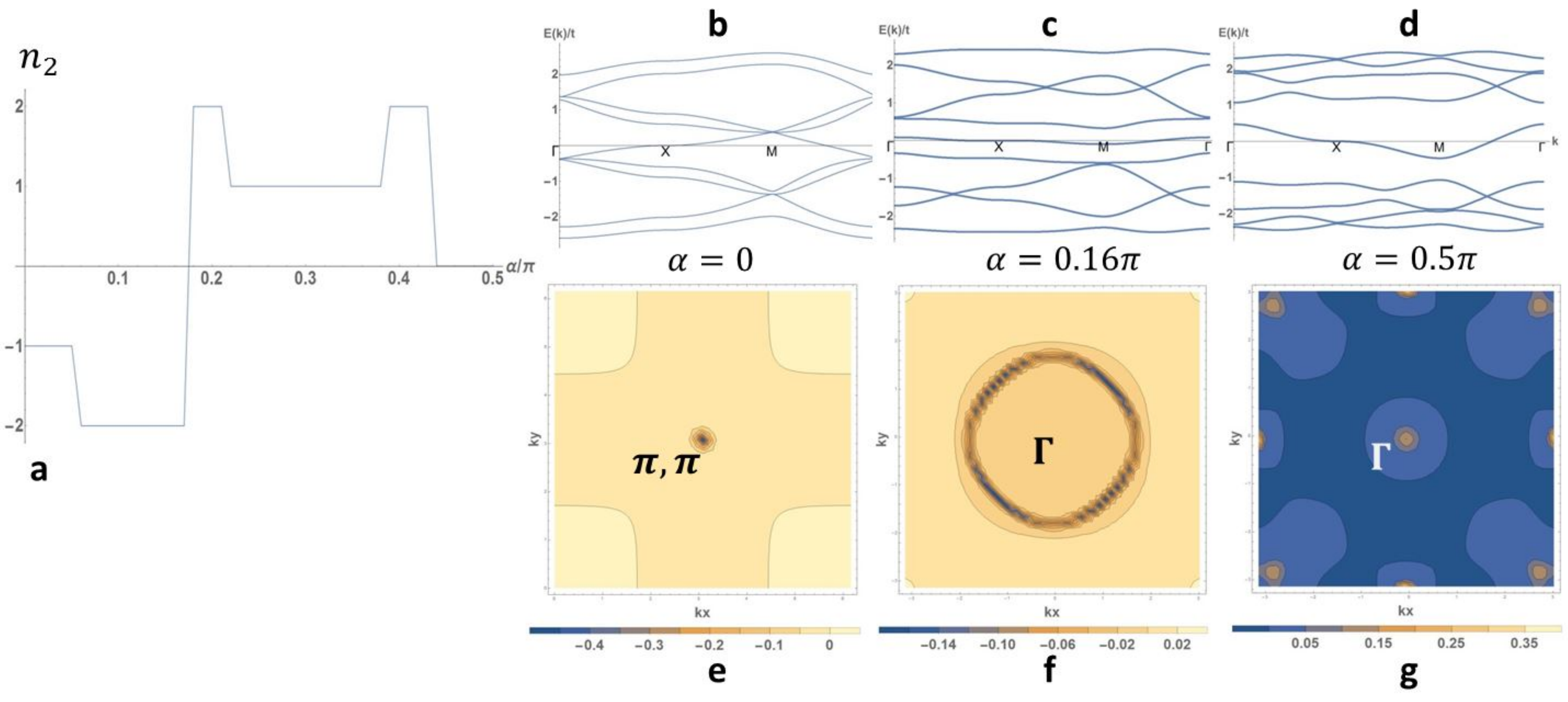}}
\caption{	
(a) Sum of Chern numbers $n_2$ as a function of $\alpha$ for the lowest two bands of a system with unit cells of 3 by 3 merons.  
These lowest two bands are well separated from the others except for $ 0.5 < \alpha < 0.67$ and $1.2 < \alpha < 1.4$ and so,
excluding these regions, the Chern number can be interpreted as the Hall conductivity.  Spin-orbit coupling 
has a non-trivial effect on the topological nature of the bands for a fixed texture.  (b,d,c) Band structure for 3 by 3 merons, 
along the principal axes of symmetry 
$\Gamma \rightarrow X \rightarrow M \rightarrow \Gamma$.
(b) $\alpha = 0$,
(c) $\alpha = 0.16 \pi$,
(d) $\alpha = 0.5 \pi$. 
(e,f,g) Berry Curvature $F_2$ for the second band of a 3 by 3 meron system for (e)$\alpha = 0$, (f) $\alpha = 0.16 \pi$ and (g) $\alpha = \pi/2$. 
Peaks in the Berry curvature identify values of $\mathbf{k}$ which contribute to the Chern number.}
\label{chernSupPanel}
\end{figure*}

In Fig. \ref{chernSupPanel}a, we show the sum of the Chern numbers for the lowest two bands
\be
n_2 = \widetilde{c}_1 + \widetilde{c}_2
\label{n2}
\ee
as a function of spin-orbit coupling calculated for a particle in a 3x3 meron texture.  In Fig. \ref{chernSupPanel}(b,c,d),
we show the energy dispersion $E(\mathbf{k})$ along the principal axes of symmetry 
%$\Gamma \rightarrow X \rightarrow M \rightarrow \Gamma$
%XXX These points are not marked-- and you don't need to say that here, just in Fig captionXXXX they are marked on axis - TM
in the Brillouin zone for various values of spin-orbit coupling.  We see that tuning
the spin-orbit coupling causes the Chern number $n_2$ to change, signaling a topological phase transition.
Each time $n_2$ changes the gap between the second band and the third band closes.  
For a range of $\alpha$ there exists a global gap between the second and third bands and in such cases the chemical potential can be placed 
within the global band gap giving $n_2$ that is proportional to the Hall conductivity
$\sigma_{xy}$.  Recent experiments have shown that the Chern number can be measured in cold atomic gases\cite{chernColdAtom} by using the transverse deflection
of an atomic cloud in reponse to an optical dipole force.  

Rather than considering only the Chern number, which gives us information about the presence of a band crossing, we can consider the momentum-resolved quantity $F_{n}$ across the Brillouin zone.
Peaks in this quantity reveal the locations of crossings of bands in momentum space. These can be experimentally detected by Aharonov-Bohm interferometry 
in momentum space.  Recent experiments\cite{bohmAharColdAtoms} have shown that the Berry phase of two Bose Einstein condensates traversing separate paths in the Brillouin zone can be measured with
high momentum resolution.  The topological phase transitions that we predict can be measured by performing momentum-space
Aharonov-Bohm interferometry of two BECs sent along 
paths in the Brillouin zone which enclose the associated Berry curvature.  

In  Fig. \ref{chernSupPanel}(e,f,g), we show the lattice Berry curvature $F_2$ of the second band for
several values of spin-orbit coupling. 
For $\alpha = 0$ (Fig \ref{chernSupPanel}e), we see one a very large negative contribution to the Berry curvature at $\mathbf{k} = (\pi,\pi)$, arising from an isolated band crossing that acts like a monopole
of Berry curvature.  For $\alpha = 0.16\pi$ (Fig \ref{chernSupPanel}f), rather than a monopole, we see a ring of Berry curvature through a portion of the Brillouin zone, indicating 
a band crossing that occurs for a locus of momenta points. When $\alpha = \pi/2$ (Fig \ref{chernSupPanel}g) we see a positive peak in the Berry curvature for $\mathbf{k} = (0,0)$ and 
four other near the corners of the Brillouin zone.  

\section{Conclusions and Discussion}

We have shown that the properties of particles moving in spin textures can be further modified by spin-orbit coupling in several important ways.  In simple textures, such as ferromagnets
and antiferromagnets, the spin-orbit coupling can be used to control the bandwidth of the particle while keeping the other properties of the band structure invariant.
We have shown that for a 3x3 meron texture the Chern number is 
nontrivial for the lowest two bands.
Above these two bands either a local gap or a global gap exists for nearly all values of $\alpha$ signaling a wide range of Chern metal and Chern insulator states.
By simply changing the value of spin-orbit coupling of the itinerant particle, the Chern number gets modified, signaling transitions between
topologically distinct quantum anomalous Hall states.

Recent work in quantum materials has focused on the current driven motion of skyrmions\cite{nagTok} and on the emergence of a finite anomalous Hall conductivity 
of $s$- electrons  coupled to chiral magnetic textures of localized $d$- electrons\cite{skxAheSpir}.  We emphasize that our model predicts a \textit{quantum} anomalous Hall effect whose Chern number is tunable by simply
changing the strength of Rashba spin-orbit coupling of the itinerant particle, a quantity which is very accessible in cold atom experiments\cite{socCold3,galSpiel}.  We have shown that 
multiple signatures of these topological phase transitions are experimentally accessible via Bohm-Aharonov interferometry and Hall drift experiments.

Rather than fixing a texture a priori, in future calculations we will
use Monte Carlo simulations to generate thermally fluctuating magnetic textures to explore the effect of temperature on the Berry curvature
and the robustness of the Chern number. Another aspect that could be important is the influence of the kinetic motion of the itinerant particle on
the magnetic texture itself. Cold atoms also provide an exceptional setting for the study of strongly correlated particles and the associated 
fractional quantum Hall physics on a lattice without 
the inclusion of an external magnetic field.
We have demonstrated that tuning between various quantum anomalous Hall states is possible by varying 
the spin-orbit coupling of the itinerant particle and it is possible that this tunability will allow for a controlled way of exploring the transitions
between various fractional quantum anomalous Hall states.

We thank W. S. Cole for useful discussions.  T.M.M. and N.T. acknowledge funding from NSF-DMR1309461.

\section{Appendix A: Details of the Projected Hopping Elements}

In this appendix, we show the explicit form of the effective hopping amplitudes $t'$ and $t''$.
In the limit of $J/t \rightarrow \infty$, the coupling 
to the texture will split the spectrum into two bands and the low energy behavior of the system will be governed entirely by the lower band which is aligned locally
with the spin texture.  Because of this local alignment, we introduce operators representing the local orientation of the spin $f_{i\sigma}^{\dagger}$ which creates a spin
aligned with the local spin $\mathbf{S}_{i}$ at site $i$.  In the usual manner, we can write these in terms of the creation and annhilation 
operators for spins aligned with the global $z$-direction.  We obtain the following unitary transformation
\be
\label{locTransAp}
  \left[ {\begin{array}{c}
   f_{j\uparrow}^{\dagger} \\
   f_{j\downarrow}^{\dagger} \\
  \end{array} } \right] =
  \left[ {\begin{array}{cc}
   g_c(\theta_j,\phi_j) & g_s(\theta_j,\phi_j) \\
   g^{*}_{s}(\theta_j,\phi_j) & -g^{*}_{c}(\theta_j,\phi_j) \\
  \end{array} } \right]%\cdot
  \left[ {\begin{array}{c}
   c_{j\uparrow}^{\dagger} \\
   c_{j\downarrow}^{\dagger} \\
  \end{array} } \right],
\ee where $g_c(\theta_j,\phi_j)$
gives the component of the projection parallel to the global axis and 
$g_s(\theta_j,\phi_j)$
gives the component of the projection antiparallel to the global axis.
The orientation of the local spin at a site $j$ is described by the polar angle $\theta_j$ and azimuthal angle 
$\phi_j$ with respect to the global $z$-axis.
In Eq. (\ref{locTransAp}) the elements of the rotation matrix are given by 
\be
g_c(\theta_j,\phi_j)=\mathrm{cos}(\theta_j / 2)\mathrm{e}^{-i\phi_j / 2}
\ee
and
\be
g_s(\theta_j,\phi_j)=\mathrm{sin}(\theta_j / 2)\mathrm{e}^{i\phi_j / 2}.
\ee
In order to derive the projected hopping Hamiltonian in Eq. (\ref{projHam}), we rewrite all operators in  Eq. (\ref{parentHam}) in terms of the
local operators given in Eq. (\ref{locTransAp}) and keep only terms of the form $f^{\dagger}_{i}f_{j}$.
We find that the term $ t_{i,i+\vec{\delta}}'$ originates from the hopping without spin-orbit coupling in the global frame and is found to be
\begin{multline}
t_{i,j}' =g_c(\theta_i,\phi_i)g^{*}_{c}(\theta_j,\phi_j)+g_s(\theta_i,\phi_i)g^{*}_{s}(\theta_j,\phi_j)\\
= \cos(\theta_i /2)\cos(\theta_{j}/2)e^{-i(\phi_i - \phi_j)/2}\\
+\sin(\theta_i /2)\sin(\theta_{j}/2)e^{i(\phi_i - \phi_j)/2}
\end{multline}
for any neighbor $j$.  The spin-orbit coupling in the global frame results in the other hopping amplitude $t_{i,i+\vec{\delta}}''$ which is found to be
%hop up
\begin{multline}
t_{i,j = i + \hat{y}}''=-i\big(g_s(\theta_i,\phi_i)g^{*}_{c}(\theta_j,\phi_j)+g_c(\theta_i,\phi_i)g^{*}_{s}(\theta_j,\phi_j) \big)\\
 =-i\big( \sin(\theta_i /2)\cos(\theta_{j}/2)e^{i(\phi_i + \phi_j)/2}\\
  +\cos(\theta_i /2)\sin(\theta_{j}/2)e^{-i(\phi_i + \phi_j)/2}\big)
\end{multline}
for hops up,
%hop down
\begin{multline}
t_{i,j = i - \hat{y}}''=i\big(g_s(\theta_i,\phi_i)g^{*}_{c}(\theta_j,\phi_j)+g_c(\theta_i,\phi_i)g^{*}_{s}(\theta_j,\phi_j) \big)\\
= i\big( \sin(\theta_i /2)\cos(\theta_{j}/2)e^{i(\phi_i + \phi_j)/2}\\
  +\cos(\theta_i /2)\sin(\theta_{j}/2)e^{-i(\phi_i + \phi_j)/2}\big)
\end{multline}
for hops down,
%hop right
\begin{multline}
t_{i,j = i + \hat{x}}''  =g_s(\theta_i,\phi_i)g^{*}_{c}(\theta_j,\phi_j)-g_c(\theta_i,\phi_i)g^{*}_{s}(\theta_j,\phi_j)\\
= \sin(\theta_i /2)\cos(\theta_{j}/2)e^{i(\phi_i + \phi_j)/2}\\
  -\cos(\theta_i /2)\sin(\theta_{j}/2)e^{-i(\phi_i + \phi_j)/2}
\end{multline}
for hops right, and
%hope left
\begin{multline}
 t_{i,j = i - \hat{x}}''=-\big(g_s(\theta_i,\phi_i)g^{*}_{c}(\theta_j,\phi_j)-g_c(\theta_i,\phi_i)g^{*}_{s}(\theta_j,\phi_j) \big) \\
= \cos(\theta_i /2)\sin(\theta_{j}/2)e^{-i(\phi_i + \phi_j)/2}\\
 -\sin(\theta_i /2)\cos(\theta_{j}/2)e^{i(\phi_i + \phi_j)/2}
\end{multline}
for hops left.

\end{document}